\documentclass[aps,pre,reprint,twocolumn,superscriptaddress]{revtex4-2}

\usepackage{amsmath}
\usepackage{amssymb}
\usepackage{graphicx}
\usepackage{bm}
\usepackage{hyperref}
\usepackage{float}

\begin{document}

\title{Detection of signals in presence of noise through Josephson junction switching currents}

\author{O. V. Pountougnigni}
\affiliation{Laboratory of Mechanics and Materials, Department of Physics, Faculty of Science, University of Yaound\'e I, Box 812, Yaound\'e, Cameroon}

\author{R. Yamapi}
\email{ryamapi@yahoo.fr}
\affiliation{Fundamental Physics Laboratory, Physics of Complex System Group, Department of Physics, Faculty of Science, University of Douala, Box 24 157 Douala, Cameroon}

\author{C. Tchawoua}
\affiliation{Laboratory of Mechanics and Materials, Department of Physics, Faculty of Science, University of Yaound\'e I, Box 812, Yaound\'e, Cameroon}

\author{V. Pierro}
\affiliation{Department of Engineering, and Gruppo Collegato Salerno INFN, Corso Garibaldi, I-82100 Benevento, Italy}

\author{G. Filatrella}
\affiliation{Department of Sciences and Technologies, and INFN, Gruppo Collegato Salerno, University of Sannio, Via F. De Sanctis, I-82100 Benevento, Italy}

\date{Received 19 February 2020; accepted 13 April 2020; published 7 May 2020}

\begin{abstract}
Josephson junctions can be employed to reveal a sinusoidal signal in presence of Gaussian noise. To mimic realistic setups, the detection is performed linearly ramping the bias current until a switch to the finite voltage occurs; the analysis of the resulting switching currents can be exploited to decide about the presence of the harmonic drive. The signal is applied in two conditions: with an unknown initial phase (incoherent strategy) and with a known initial phase (coherent strategy). In both conditions, the analysis of the efficiency of the detection, performed through the signal-to-noise ratio, as estimated by the Kumar-Carrol index, shows that the dependence upon the Josephson junction ramp rate is beneficial, especially for relatively fast speed. One can conclude that the collection of the switching currents is a robust technique, and thus it is possible to exploit the advantages of a predetermined finite time to collect the data.
\end{abstract}

\doi{10.1103/PhysRevE.101.052205}

\maketitle

\section{Introduction}
The detection of a sinusoidal signal is a widespread problem in basic physics and applications. An example is threshold detection [1--3], based on the possibility of ascertaining the presence of a signal via the transition from one metastable state to another. For instance, Josephson junction (JJ) devices can move, under the effect of both the applied external signal and the unavoidable fluctuations from the metastable superconducting state to a finite voltage state; the analysis of the passages can be employed to infer the presence of a perturbation. The technique is similar to the one exploited since the early appearance of the mechanism of stochastic resonance [4] for the detection of digital subthreshold signals with Schmitt triggers [5,6]. More recently, it has been suggested, through simulations of neuronal models, that noise can be beneficial for signal detection [7], also in the presence of correlated noise [8].

There are some advantages in choosing JJs: (1) as superconducting elements JJs can operate at very low temperature (close to absolute zero or to the quantum limit [9]) and are therefore possibly affected by low intrinsic thermal noise and (2) JJs are extremely fast elements [10], as fast as a few hundred GHz or even close to the THz region [11]. Since the pioneering works [12,13] many experiments have been carried out to highlight the role of noise in ac-driven JJs, for instance, to pinpoint the quantum behavior [14,15], to reach the single photon limit [16--18], or to detect the phase of an applied signal [19]. More recently, JJs as threshold detectors have been characterized through the analysis of the moments of the distribution of the events [20--22].

Since the pioneering experiments [13], the switching time distribution has proven very sensitive to the signal amplitude, and it is conceivable that one can collect the switching times to discriminate between two situations: (1) the exit is caused by pure noise (no signal is present), and (2) exit is caused by the combined action of noise and a sinusoidal excitation (the signal is present). In the theory of signal detection, the quantitative analysis is evaluated through the determination of the detector performance. A common tool for the estimate of the signal-to-noise ratio (SNR) of the detector is the Kumar-Carrol (K-C) index [23--25]. The index is a suitable indicator of the performances of the detector and allows one to determine the most favorable condition in terms of the applied frequency and bias current and of the device dissipation [25]. The performances, however, also depend on the adapted strategy, as the quantity that can be monitored varies from the sample mean to a more refined likelihood ratio test [26]. It is therefore natural to look for the performances in a widely employed setup, the case in which the bias current depends upon the time [27]. As the performances of the detector depend on the conditions in which the signal is applied, one can further consider two strategies: coherent (a prescribed phase relation between the applied signal and the current ramp) and incoherent (an unknown phase between the two drivers) [24].

In brief, the aim of the present work is to study the effect of time-dependent bias on JJs, employed as detectors, analyzing the performances through the K-C index. To do so, we proceed as follows. In Sec.~II we describe the JJ model, modified to accommodate for the signal effect and the bias scheme. In Sec.~III the methods to analyze the switching events is described, and in Sec.~IV are collected the resulting performances of the sketched device. Section~V summarizes the main findings and the outlook.

\section{Model and Simulations Method}
In this section the model for the JJ dynamics, as modified by a sinusoidal signal and Gaussian noise, is laid down; furthermore, the numerical method employed in the simulations is also outlined.

\subsection{Electrical model}
The electrical model we deal with in the present analysis, displayed in Fig.~1, is the resistively shunted and capacitively junction model [10]. It consists of a capacitor $C_J$ and a parallel connected resistor $R_J$, while J in the rectangle represents the ideal Josephson junction element ($I_0$ is the maximum Josephson current that describes the nonlinear relation between the current and the gauge invariant phase difference $\varphi = \varphi_1 - \varphi_2$ across the two superconductors):
\begin{equation}
I_J = I_0 \sin \varphi.
\end{equation}
The Josephson element constitutive voltage equation reads
\begin{equation}
V = \frac{\hbar}{2e} \frac{d\varphi}{dt}.
\end{equation}

\begin{figure}[htbp]
\centering
\includegraphics[width=\columnwidth]{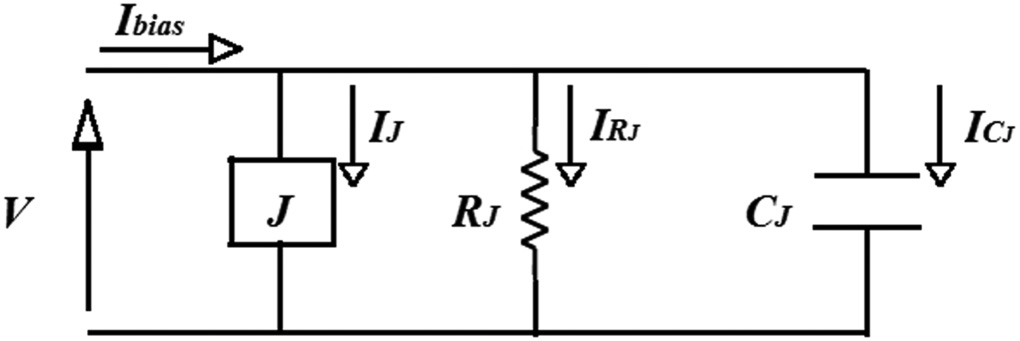}
\caption{Scheme of the electrical model of a Josephson junction element.}
\label{fig:fig1}
\end{figure}

The current bias consists of a continuous dc-current generator $I_b$, and the signal is modeled through an alternating (ac) term: $I_{\text{ac}} \sin (\Omega t + \theta)$, disturbed by a noise current $I_n$. A sinusoidal term is added on the top of noise, as well as to let the quantum behavior emerge [28,29] or to investigate the dependence upon the initial conditions [30]. Thus, denoting with $I_J$ the current through the Josephson element, $I_{R_J}$ the current through the JJ resistor, and $I_{C_J}$ the current through the junction capacitance, the current balance reads
\begin{equation}
I_{C_J} + I_{R_J} + I_J = I_b + I_{\text{ac}} \sin (\Omega t + \theta) + I_n.
\end{equation}
If we set
\begin{equation}
I_{C_J} = C_J \frac{dV}{dt} = C_J \frac{\hbar}{2e} \frac{d^2\varphi}{dt^2}, \quad I_{R_J} = \frac{V}{R_J} = \frac{\hbar}{R_J 2e} \frac{d\varphi}{dt},
\end{equation}
the model is described by the following second-order differential equation:
\begin{equation}
C_J \frac{\hbar}{2e} \frac{d^2\varphi}{dt^2} + \frac{\hbar}{R_J 2e} \frac{d\varphi}{dt} + I_0 \sin \varphi = I_b + I_{\text{ac}} \sin (\Omega t + \theta) + I_n,
\end{equation}
which is a Langevin equation [31]. Introducing the Josephson frequency $\omega_J = \sqrt{2eI_0/C_J\hbar}$, Eq.~(5) can be cast in time-normalized units ($\tau = \omega_J t$), as follows:
\begin{equation}
\frac{d^2\varphi}{d\tau^2} + \alpha \frac{d\varphi}{d\tau} + \sin \varphi = \gamma + \gamma_{\text{ac}} \sin (\omega\tau + \theta) + \zeta (\tau),
\end{equation}
where the parameters are defined as
\begin{equation}
\alpha = \frac{1}{R_J C_J \omega_J}, \quad \gamma = \frac{I_b}{I_0}, \quad \gamma_{\text{ac}} = \frac{I_{\text{ac}}}{I_0}, \quad \omega = \frac{\Omega}{\omega_J}.
\end{equation}

\begin{figure}[htbp]
\centering
\includegraphics[width=\columnwidth]{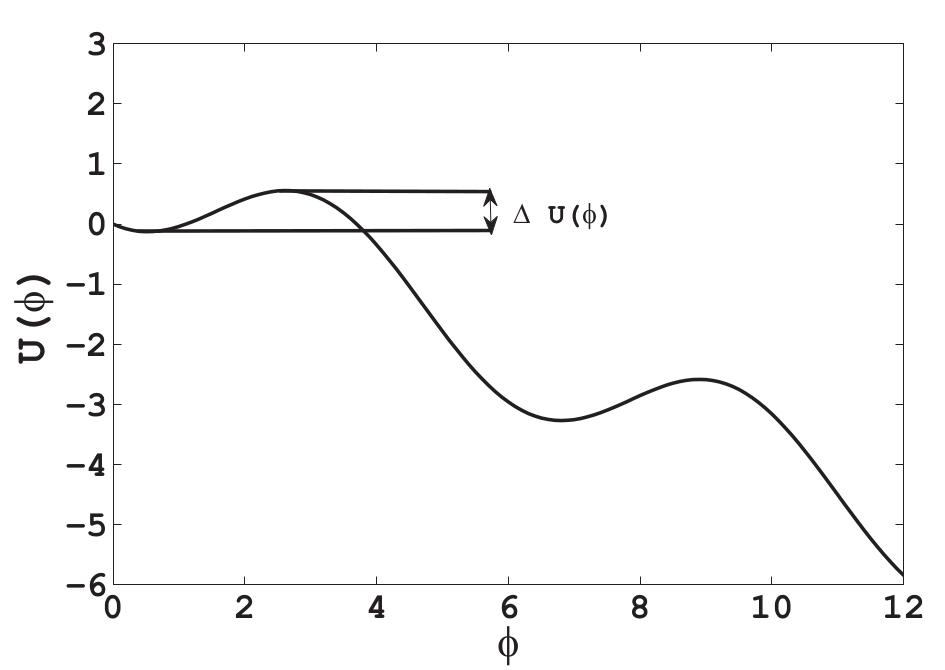}
\caption{The washboard potential for $\gamma = 0.5$. The energy barrier $\Delta U$ is given by Eq.~(10).}
\label{fig:fig2}
\end{figure}

The statistical features of the noisy term $\zeta (\tau)$ are determined by
\begin{equation}
\langle \zeta (\tau) \rangle = 0, \quad \langle \zeta (\tau)\zeta (\tau') \rangle = 4D\delta(\tau - \tau'),
\end{equation}
where $D = k_B T \omega_J / (R I_0^2)$ is the normalized noise intensity ($k_B$ is the Boltzmann constant and $T$ the absolute temperature).

A washboard potential is associated with Eq.~(12) [10,31]:
\begin{equation}
U(\varphi) = -\gamma\varphi + [1 - \cos(\varphi)].
\end{equation}
For $\gamma < 1$, the potential has metastable wells with a barrier height [31]:
\begin{equation}
\Delta U(\gamma) = 2 \left[ \sqrt{1 - \gamma^2} - \gamma \cos^{-1}(\gamma) \right].
\end{equation}
The dependence of the potential $\Delta U$ versus the normalized current $\gamma$ is shown in Fig.~2. For a constant energy barrier the resulting escape rate $\Gamma(\gamma)$ is given by the Kramers approximation:
\begin{equation}
\Gamma(\gamma) = (1 - \gamma^2)^{1/4} e^{-\frac{\Delta U(\gamma)}{D}}.
\end{equation}
The prefactor depends upon the dissipation and the approximation employed [32]. In Eq.~(11) is displayed a major contribution arising from the bias-dependent resonant mode [10].

Here one is interested to ascertain whether the detection of the sinusoidal term is enhanced if the external bias current includes a time-dependent term; consequently Eq.~(6) becomes
\begin{equation}
\frac{d^2\varphi}{d\tau^2} + \alpha \frac{d\varphi}{d\tau} + \sin \varphi = \gamma (\tau) + \gamma_{\text{ac}} \sin (\omega\tau + \theta) + \zeta (\tau).
\end{equation}
Where the dc current is not constant anymore, for the bias $\gamma$ is the ramp:
\begin{equation}
\gamma (\tau) = v\tau.
\end{equation}
Here $v$ is the speed at which the bias rises is swept from zero to the critical current; that is, in a time $1/v$ the bias reaches the normalized critical current $\gamma = 1$. At this current the JJ, in the absence of the signal and in the noiseless case ($\gamma_{\text{ac}} = 0$, $D = 0$) switches to a finite voltage $d\varphi/d\tau \neq 0$, and the current in correspondence of the event can be measured [33]. The ramp speed is normalized to the JJ characteristic time, which is in the order of 100 GHz, and therefore one can assume that in normalized units it should read at least $10^{-3}$ and that can be easily as slow as $10^{-6}$. The bias ramp is convenient, for it guarantees that the switching events are collected in a predetermined time: at least a switch event occurs in a time $1/v$. It is therefore very common to encounter the data in the form of switching current distributions [34]. Moreover, the bias changes the potential; therefore the system explores a range of trapping energies, a feature that has proved convenient in the analysis of non-Gaussian, L\'evy-type, noise sources [35,36]. Finally, the relation (13) converts the switching time $\tau_{\text{sw}}$ in a switching current through the relation
\begin{equation}
\gamma_{\text{sw}} = v\tau_{\text{sw}}.
\end{equation}
Given the simple relation between time and switching currents, in this work the terms ``switching times'' and ``switching currents'' will be used as interchangeable words.

The switching currents' distribution $P(\gamma)$ obtained with a bias ramp (13) and the escape rate $\Gamma(\gamma)$ from the static potential (10) are connected by the relation [37]
\begin{equation}
P(\gamma) = \Gamma(\gamma)\left(\frac{d\gamma}{d\tau}\right)^{-1}\left[1 - \int_0^\gamma P(u)\,du\right].
\end{equation}
The above relation between the escape rate $\Gamma(\gamma)$ and the switching current distribution $P(\gamma)$ highlights that the switching current distribution depends upon the escape rates of all barrier heights $0 \le \Delta U \le 2$. Because of the intricate relation between the properties of the escape at constant bias (11) and the distribution of the switching currents (15), it is not guaranteed that the detection performances are the same at constant bias in respect to the bias ramp; hence the call for a detailed analysis of the performances if the bias is swept. The results of the detection properties of the switching currents are the main objective of this paper and will be discussed in Sec.~IV.

\subsection{Algorithms for numerical simulations}
For stochastic simulations of the differential equation (12) we use the Euler algorithm [38], and the Box-Muller method [39] is employed to generate Gaussian white noise from two random numbers $a$ and $b$ which are uniformly distributed in the unit interval $[0, 1]$:
\begin{equation}
a = \text{random number}, \quad b = \text{random number},
\end{equation}
\begin{equation*}
\zeta_n^i = \sqrt{-4D\Delta\tau \log(a)}\cos(2\pi b).
\end{equation*}
The time step through all the simulation is $\Delta\tau = 0.0001$. The stochastic results are averaged over as many realizations as are necessary to obtain reliable results, generally $10^4$. Two parameters that appear in Eq.~(6), $\alpha$ and $D$, have been set to a constant value through all simulations, namely, $\alpha = 0.2$ and $D = 0.05$ to thoroughly explore the effects of the other parameters (the speed $v$, the signal amplitude $\gamma_{\text{ac}}$, and the frequency $\omega$). For fast JJ (a frequency around 100 GHz) and realistic values of the temperature (around 1K to be below the low $T_c$ transition temperature), the normalized parameters here chosen ($\alpha = 0.2$ and $D = 0.05$) entail very demanding values of the critical current (about $0.1\ \mu\text{A}$), the shunt resistor (around $10\text{ k}\Omega$), and of the capacitance (around $2\text{ fF}$). The critical current and the capacitance would increase, while keeping the other parameters constant, decreasing both the normalized dissipation $\alpha$ and noise intensity $D$ to explore a much wider range, possibly with parallel CUDA [40]. For instance, increasing by an order of magnitude the critical current and the capacitance would result in a decrease of an order of magnitude of the dissipation parameter ($\alpha \approx 0.02$), and of two orders of magnitude of the noise correlator ($D \approx 0.0005$).

\section{Detection Strategies}
In the context of signal detection the phase $\theta$ of the signal in Eq.~(12) might be known or not; in the following are described the two corresponding procedures.

Where the phase of the applied signal is unknown, one can mimic the ignorance of the initial conditions applying a signal with a random initial phase (incoherent detection). Physically, it corresponds to apply the signal to the junction with an unknown phase $\theta$, for instance, if the junction is exposed to an uncontrolled source of radiation. Such blindness of the detection system can be modeled through an initial phase that reads
\begin{equation}
\theta \in [0, 2\pi[,
\end{equation}
with uniform distribution. When the JJ switches to the running, finite voltage state, the switching time $\tau_i$ is recorded, the system is reset to the static state, and the signal is applied again adding $\omega\tau_i$ to the previous initial phase.

Alternatively, the second strategy is called coherent, for the sinusoidal signal is always applied with the same null initial phase, for instance, $\theta = 0$. It is known that the initial phase does have an impact on the distribution times [24,25,30], therefore the case of zero initial phase is but an example. When the switch occurs, the switching time is recorded and the system is shielded from the signal. Then the system is reset to the static state with the same initial phase.

Repeating the procedure $N$ times, one retrieves a collection of switching times [24,25]:
\begin{equation}
\bm{\tau} = \{\tau_i\}_{i=1}^N.
\end{equation}
The collected data are to be analyzed to decide whether or not a sinusoidal excitation is embedded into noise. The performances of the detection can be summarized by the index $d_{\text{KC}}$ [23], such as
\begin{equation}
d_{\text{KC}} = \frac{|\langle \tau_{\text{sw}} \rangle_{\text{sig}} - \langle \tau_{\text{sw}} \rangle_{\text{noi}}|}{\sqrt{\frac{1}{2}\left(\sigma^2(\tau_{\text{sw}})_{\text{sig}} + \sigma^2(\tau_{\text{sw}})_{\text{noi}}\right)}},
\end{equation}
where $\langle \tau_{\text{sw}} \rangle_{\text{sig}}$ denotes the estimated average switching time in the presence of the signal and the estimate of the corresponding standard deviation $\sigma (\tau_{\text{sw}})_{\text{sig}}$. Analogously, $\langle \tau_{\text{sw}} \rangle_{\text{noi}}$ is the estimated average switching time without signal, and $\sigma (\tau_{\text{sw}})_{\text{noi}}$ the corresponding estimated standard deviation (all averages are taken over noise realizations). In the case of a Gaussian distribution---which is not exactly the case in the present collection of switching currents---the $d_{\text{KC}}$ corresponds to the signal-to-noise ratio (SNR). In this work the index $d_{\text{KC}}$ (19) is employed as an estimate, or a proxy, of the SNR.

\section{Results and Discussion of the Detection Performances}
In the following section, the variable bias technique is characterized for the detection of a sinusoidal drive. Specifically, the performances are analyzed as a function of the intensity of signal and of the bias ramp speed $v$, for both the incoherent and coherent strategy. The detector is analyzed through the response of the average switching time and of an estimate of the SNR.

\subsection{The complementary cumulative distribution function of the escape time}
To emphasize the effect of the signal on the escape times, it is convenient to introduce the complementary cumulative distribution function $\text{CCDF}(X)$. For a random current that depends upon a coordinate $x$, the CCDF shows the probability that the phenomenon occurs for $x > X$, which is complementary to the cumulative distribution function $\text{CDF}(X)$, which represents the probability that the phenomenon occurs for $x < X$. 

\begin{figure*}[t]
\centering
\includegraphics[width=2\columnwidth]{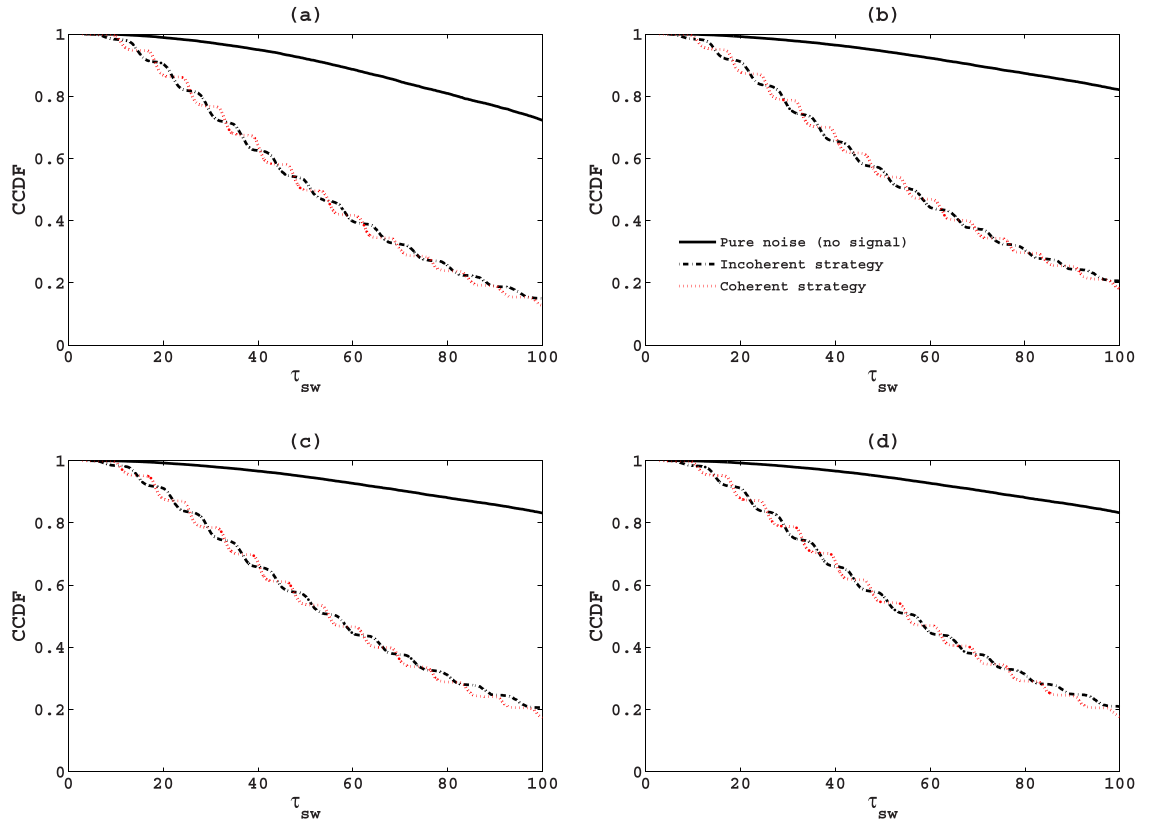}
\caption{Complementary cumulative distribution function, CCDF, of the escape time for pure noise (solid line) and for a combination of noise with a sinusoidal signal ($\gamma_{\text{ac}} = 0.2$) in the incoherent (dot-dashed line) and coherent (dashed red line) cases. Parameters of simulation are $D = 0.05$, $\omega = 0.86$, $\alpha = 0.2$ with (a) $v = 10^{-3}$; (b) $v = 10^{-4}$; (c) $v = 10^{-5}$; (d) $v = 10^{-6}$.}
\label{fig:fig3}
\end{figure*}

Thus, Fig.~3 shows the CCDF of the escape times in the presence of the sinusoidal signal ($\gamma_{\text{ac}} = 0.2$, broken line) and without a signal ($\gamma_{\text{ac}} = 0$, solid line) obtained simulating Eq.~(12), for several different values of the speed $v$. The four panels, each corresponding to a velocity $v$ of the current bias ramp $\gamma (\tau)$, visualize the effect of the ramp speed. The difference between the CCDF with and without the signal is a clear indication that JJ response to a drive is not much influenced by the ramp velocity $v$ and can therefore be employed for signal detection. However, it is relevant for signal detection that the CCDFs in the presence of the sinusoidal signal for the two strategies are similar, with a little difference in slope.

\subsection{Results for the incoherent strategy}
Figure~4 displays for the incoherent strategy the average switching time $\langle \tau_{\text{sw}} \rangle$ as a function of the frequency $\omega$ of the sinusoidal signal for different values of signal intensity $\gamma_{\text{ac}}$. The curves exhibit the phenomenon of resonance at the frequency $\omega_0 \approx 0.80$ [24]. It is also to be noticed that the frequency at which the resonance occurs weakly depends upon the amplitude of the signal, as expected for a nonlinear oscillator. Therefore, the best conditions to detect a signal weakly also depend upon the amplitude of the signal itself.

\begin{figure}[htbp]
\centering
\includegraphics[width=\columnwidth]{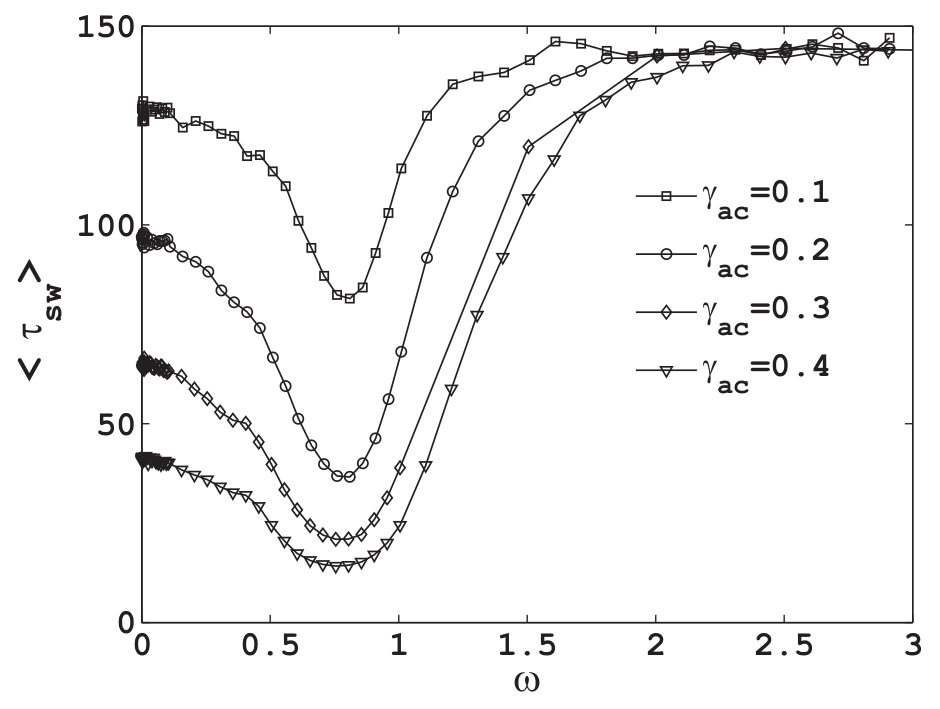}
\caption{Average escape time $\langle \tau_{\text{sw}} \rangle$ as a function of the driving frequency $\omega$ for the incoherent strategy, to highlight the different amplitudes of the signal. The symbols denote different amplitudes: square: $\gamma_{\text{ac}} = 0.1$, circle: $\gamma_{\text{ac}} = 0.2$, diamond: $\gamma_{\text{ac}} = 0.3$, down triangle: $\gamma_{\text{ac}} = 0.4$. Parameters of simulations are $D = 0.05$, $v = 10^{-6}$, $\alpha = 0.2$.}
\label{fig:fig4}
\end{figure}

Moreover, in Fig.~4 it is evident that as the signal intensity increases the minimum of the average switching time decreases. As the signal frequency approaches $\omega \approx 2.0$, the asymptotic behavior of different curves is approximately independent of the signal intensity.

The response of a JJ, in the form of a change of the average escape time, is therefore a qualitative indication that JJ can be employed as detectors. The performances of the detection can be characterized more accurately through the SNR as estimated by the K-C index [23]. This is shown in Fig.~5 for several different values of signal intensity. At the resonant frequency the K-C index increases with the the signal amplitude. For example, one reads $d_{\text{kc}} \approx 0.8$ for $\gamma_{\text{ac}} = 0.1$, while $d_{\text{kc}}$ rises to $\approx 1.93$ for $\gamma_{\text{ac}} = 0.4$. In the region at higher frequency, $\omega \gtrsim 2$, the average switching time is almost independent of the signal amplitude, and in fact the index $d_{\text{kc}}$ vanishes, which tells us that this region is not suitable for detection.

\begin{figure}[htbp]
\centering
\includegraphics[width=\columnwidth]{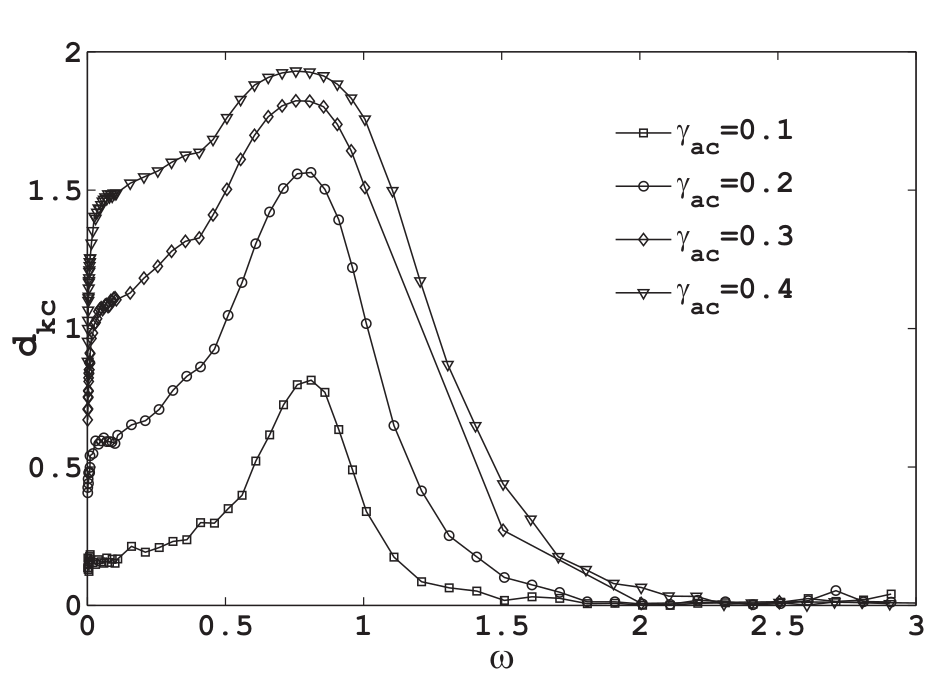}
\caption{K-C index $d_{\text{KC}}$ as a function of the driving frequency $\omega$ for the incoherent strategy for several values of the sinusoidal amplitude $\gamma_{\text{ac}}$. The symbols denote different amplitudes: square: $\gamma_{\text{ac}} = 0.1$, circle: $\gamma_{\text{ac}} = 0.2$, diamond: $\gamma_{\text{ac}} = 0.3$, down triangle: $\gamma_{\text{ac}} = 0.4$. Parameters of simulations are $D = 0.05$, $v = 10^{-6}$, $\alpha = 0.2$.}
\label{fig:fig5}
\end{figure}

The effect of the bias current ramp speed is reported in Figs.~6 and 7. For $v = 10^{-3}$ the detector's best performances read $d_{\text{kc}} = 1.76$ at the resonant state. Still better performances are expected for $v = 10^{-2}$, as the deviations of the average escape time are still more marked; see Fig.~6. We note that some data in correspondence of $v = 10^{-2}$ are missed, for the speed is so high that escape occurs when the ramp is almost completed, in the proximity of the critical current (it actually coincides with the critical current, within numerical accuracy). However, it is evident that the detector's performances are enhanced for fast ramp speed. The sweep bias speed is also related to the observation time $1/v$ necessary to retrieve a single escape, which obviously shortens for large speed. In conclusion, the results indicate that it is favorable to use a large ramp speed, for it results in two positive effects: it increases the SNR (as estimated through the K-C index) and shortens the observation time.

\begin{figure}[htbp]
\centering
\includegraphics[width=\columnwidth]{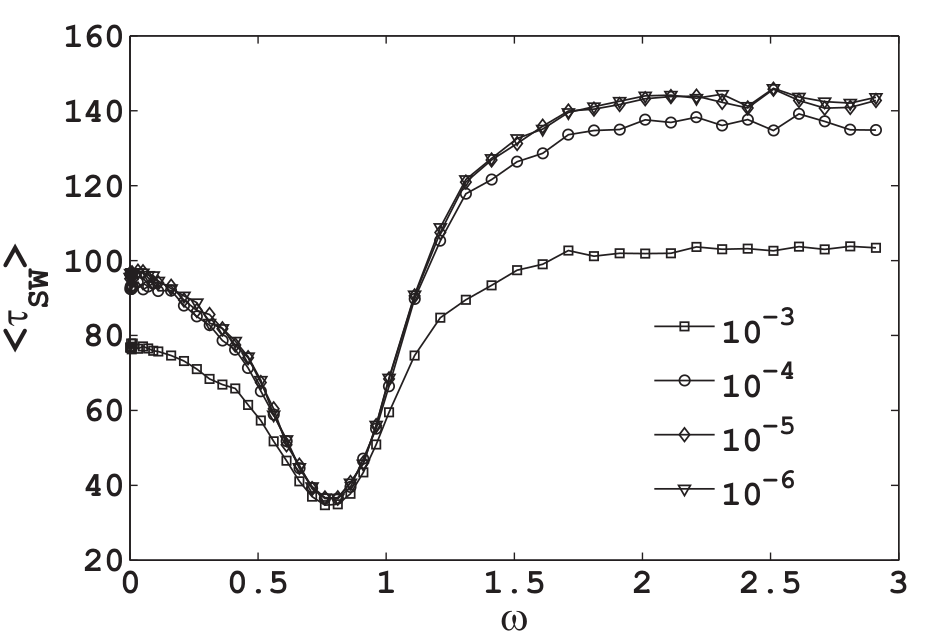}
\caption{Average escape time $\langle \tau_{\text{sw}} \rangle$ as a function of the driving frequency $\omega$ for the incoherent strategy, to highlight the different amplitudes of the signal. The markers for different curves are star: $v = 10^{-2}$, square: $v = 10^{-3}$, circle: $10^{-4}$, diamond: $10^{-5}$, down triangle: $10^{-6}$. Parameters of the simulations are $D = 0.05$, $\gamma_{\text{ac}} = 0.2$, $\alpha = 0.2$.}
\label{fig:fig6}
\end{figure}

\begin{figure}[htbp]
\centering
\includegraphics[width=\columnwidth]{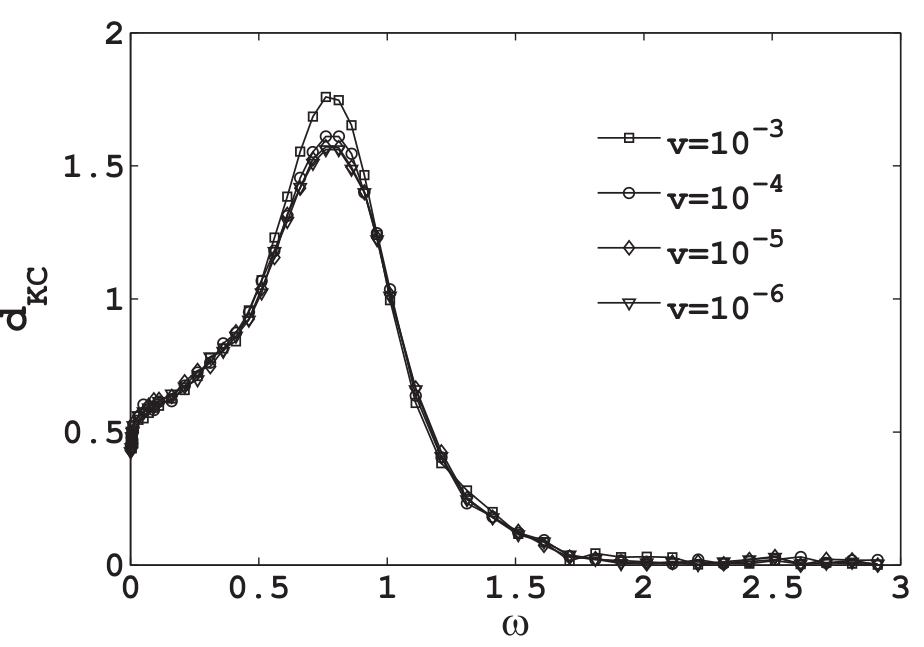}
\caption{K-C index $d_{\text{KC}}$ as a function of the driving frequency $\omega$ for the incoherent strategy, to highlight the different amplitudes of the signal. The markers for different curves are square: $v = 10^{-3}$, circle: $10^{-4}$, diamond: $10^{-5}$, down triangle: $10^{-6}$. Parameters of the simulations are $D = 0.05$, $\gamma_{\text{ac}} = 0.2$, $\alpha = 0.2$.}
\label{fig:fig7}
\end{figure}

\subsection{Results for the coherent strategy}
In the coherent strategy case each sequence begins with the same initial phase $\theta = 0$. The analysis of the average switching time is reported in Fig.~8, which displays a dip at low frequencies which is not present in the incoherent strategy. The behavior of the dip amplitude as a function of the signal frequency is also reported in Table~I.

\begin{figure}[htbp]
\centering
\includegraphics[width=\columnwidth]{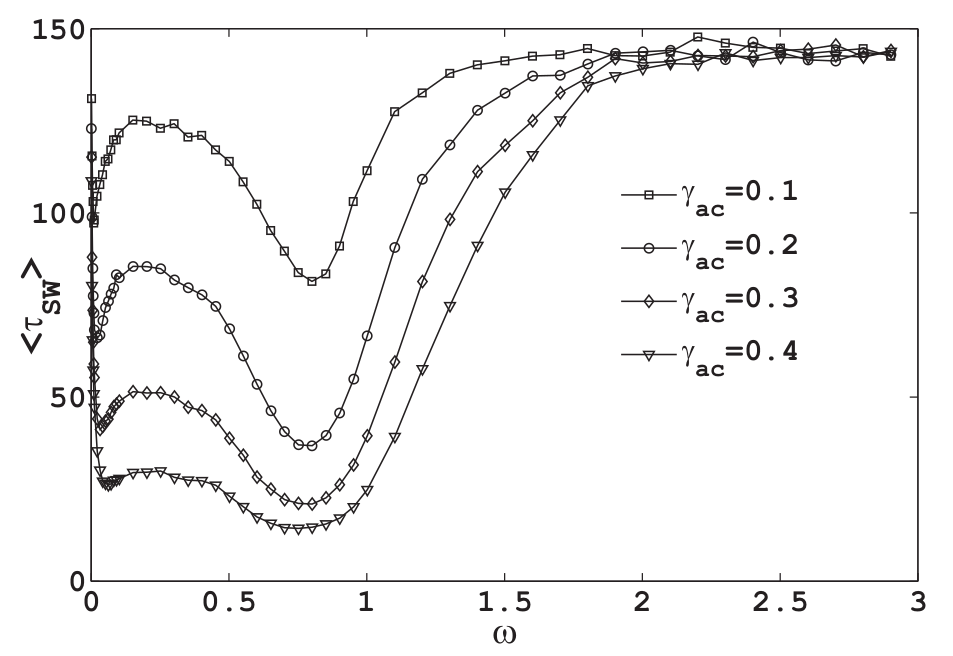}
\caption{Average escape time $\langle \tau_{\text{sw}} \rangle$ as a function of the driving frequency $\omega$ for the coherent strategy to highlight the different amplitudes of the signal. The markers for different curves are square: $\gamma_{\text{ac}} = 0.1$, circle: $\gamma_{\text{ac}} = 0.2$, diamond: $\gamma_{\text{ac}} = 0.3$, down triangle: $\gamma_{\text{ac}} = 0.4$. Parameters of the simulations are $D = 0.05$, $v = 10^{-6}$, $\alpha = 0.2$.}
\label{fig:fig8}
\end{figure}

As expected, the performance of the detection reveals that the SNR increases with the amplitude of the signal (see Fig.~9 and Table~I). There are, however, only marginal differences with the incoherent strategy; for example, in both cases it is observed the phenomenon of noise-independent resonance at the frequency $\omega_0 \approx 0.80$ and the asymptotic behavior of different curves when the drive frequency approaches 2.

\begin{table}[H]
\caption{Frequency at which a dip is observed as a function of the signal amplitude $\gamma_{\text{ac}}$ for the coherent detection strategy. The last column displays the SNR of the signal, as estimated through the K-C index.}
\label{tab:tab1}
\centering
\renewcommand{\arraystretch}{1.35}
\begin{tabular*}{\linewidth}{@{\extracolsep{\fill}}ccc}
\hline\hline
Amplitude $\gamma_{\text{ac}}$ & Frequency $\omega$ & K-C index $d_{\text{KC}}$ \\
\hline
0.1 & 0.009 & 0.586 \\
0.2 & 0.012 & 1.122 \\
0.3 & 0.032 & 1.468 \\
0.4 & 0.052 & 1.722 \\
\hline\hline
\end{tabular*}
\end{table}

\begin{figure}[htbp]
\centering
\includegraphics[width=\columnwidth]{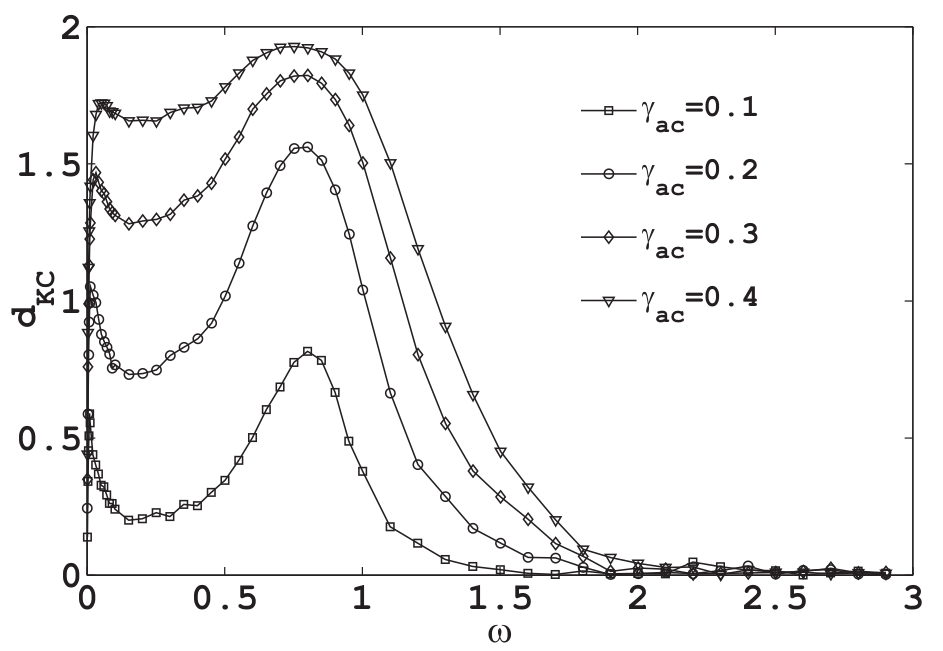}
\caption{K-C index $d_{\text{KC}}$ as a function of the driving frequency $\omega$ for the coherent strategy, to highlight the different amplitudes of the signal. The markers for different curves are square: $\gamma_{\text{ac}} = 0.1$, circle: $\gamma_{\text{ac}} = 0.2$, diamond: $\gamma_{\text{ac}} = 0.3$, down triangle: $\gamma_{\text{ac}} = 0.4$. Parameters of the simulations are $D = 0.05$, $v = 10^{-6}$, $\alpha = 0.2$.}
\label{fig:fig9}
\end{figure}

The effect of the ramp speed is reported in Figs.~10 and 11. Also the response to the bias ramp speed is similar to the incoherent strategy of Sec.~IV~C. As expected, in this case as well it is not possible to explore the whole parameter region for a very high speed, $v = 10^{-2}$, inasmuch as the escapes occur too close to the critical current at $\omega > 0.9$. However, the analysis shows an interesting region, reported in Table~II, which does not exist for the incoherent strategy. Table~II confirms that upon increasing the ramp speed the K-C index improves, or at least it stays constant, and therefore it can be concluded that a high ramp speed is favorable for signal detection also in the case of a coherent strategy.

\begin{figure}[htbp]
\centering
\includegraphics[width=\columnwidth]{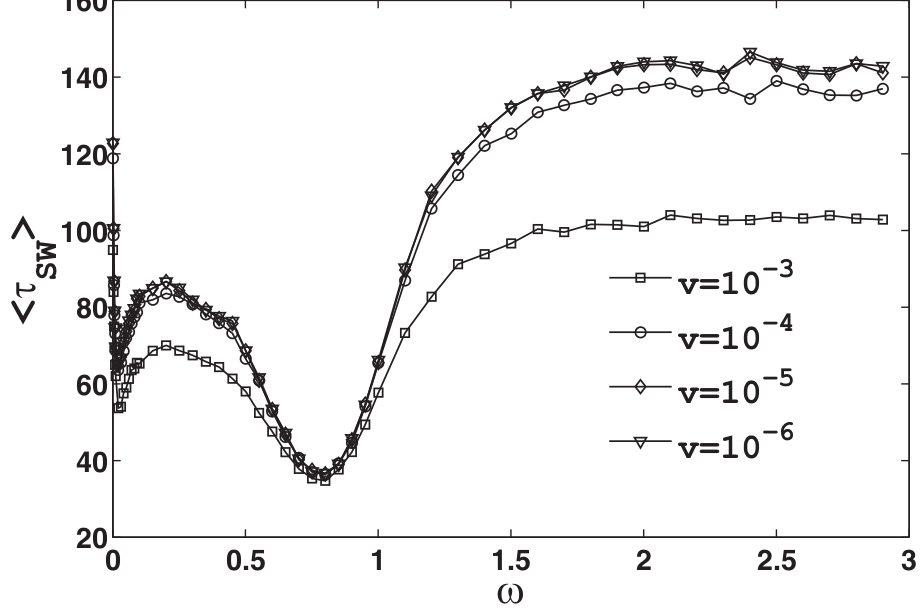}
\caption{Average escape time $\langle \tau_{\text{sw}} \rangle$ as a function of the driving frequency $\omega$ for the coherent strategy to highlight the effect of different bias ramp speed. The markers for different curves are star: $v = 10^{-2}$, square: $v = 10^{-3}$, circle: $10^{-4}$, diamond: $10^{-5}$, down triangle: $10^{-6}$. Parameters of the simulations are $D = 0.05$, $\gamma_{\text{ac}} = 0.2$, $\alpha = 0.2$.}
\label{fig:fig10}
\end{figure}

\begin{figure}[htbp]
\centering
\includegraphics[width=\columnwidth]{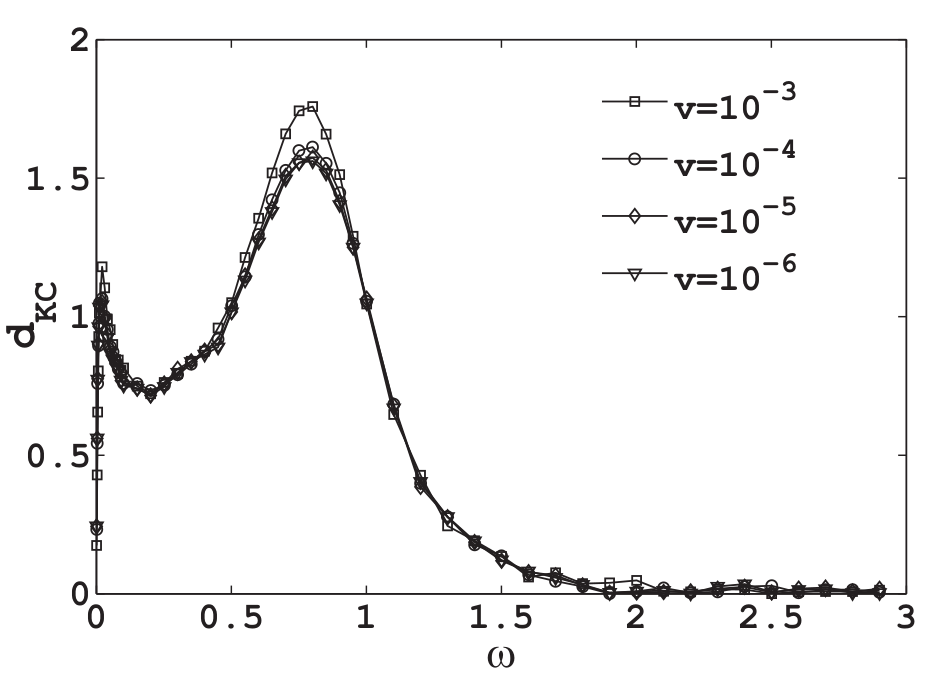}
\caption{K-C index $d_{\text{KC}}$ as the function of the driving frequency $\omega$ for the coherent strategy to highlight the effect of different bias ramp speed. The markers for different curves are square: $v = 10^{-3}$, circle: $10^{-4}$, diamond: $10^{-5}$, down triangle: $10^{-6}$. Parameters of the simulations are: $D = 0.05$, $v = 10^{-6}$, $\alpha = 0.2$.}
\label{fig:fig11}
\end{figure}

\begin{table}[H]
\caption{Dependence of the frequency at which a dip is observed as a function of the bias ramp speed $v$. The last column displays the SNR of the signal, as estimated through the K-C index for the amplitude $\gamma_{\text{ac}} = 0.2$}
\label{tab:tab2}
\centering
\renewcommand{\arraystretch}{1.35}
\begin{tabular*}{\linewidth}{@{\extracolsep{\fill}}cccc}
\hline\hline
Ramp speed & Frequency & Average switching & K-C index \\
$v$ & $\omega$ & time $\langle \tau_{\text{sw}} \rangle$ & $d_{\text{KC}}$ \\
\hline
$10^{-3}$ & 0.021 & 53.65 & 1.180 \\
$10^{-4}$ & 0.021 & 63.63 & 1.065 \\
$10^{-5}$ & 0.021 & 64.78 & 1.042 \\
$10^{-6}$ & 0.021 & 65.27 & 1.042 \\
\hline\hline
\end{tabular*}
\end{table}

\section{Conclusions}
In summary, the detection of sinusoidal signals using a nonlinear device, an underdamped Josephson junction, has been analyzed considering a ramp of the dc bias current at velocity $v$. The resulting distribution of the switching currents has been characterized as a function of the bias ramp speed in terms of the SNR to ascertain if it is suitable for effective detection. The underlying reason is that a nonconstant bias is convenient, for it guarantees that an escape is recorded each cycle of the bias current, thus securing the number of collected data per unit time of detection. However, two natural questions arise. First, do the switching times so collected allow us to detect the difference between the case of pure noise from the case in which a sinusoidal term is present? To this question the answer is positive, for we have confirmed that the average switching current [or the average escape time; as per Eq.~(13) the two quantities are simply proportional] is indeed sensitive to an oscillatory term. The second question is how do we show the influence of the bias sweep velocity on the efficacy of the detection? Also in this case the answer is positive---actually twofold positive: the detection is possible, and it improves in the most favorable condition of large speed, in which the data are collected more quickly. One concludes that it is convenient to use the ramp current setting, and to employ as high of a value of the ramp speed as electronics allows. In fact, the speed of the bias ramp does not influence the performance at low speed and improves the performances when the speed becomes high, $v \approx 0.001$ (in normalized units). The point is quite delicate, for a normalized speed of 0.001 or above easily entails an actual bias frequency of several hundreds of MHz, which is not easy to reach in this context.

In more detail the findings are the following. Using two strategies [24], coherent and incoherent, to determine the statistical features of the detector through the Kumar-Carrol index $d_{\text{KC}}$ [23], it is possible to determine the performances of JJ to reveal sinusoidal signals in the presence of noise. For both strategies the JJ appears to be very sensitive to the signal in proximity of the junction resonance. The dynamics shows an asymmetric behavior: relatively good performance for low frequencies and very low sensitivity above the resonance [24]. Also, the change in the amplitude of the signal weakly changes the resonant frequency and induces a significant variation of the average switching time (see Figs.~4 and 8). Finally, the change in the bias ramp speed changes neither the resonant frequency nor the average switching time at the resonance (see Figs.~6 and 10).

To wrap up in a sentence the main finding: it is definitively beneficial for signal detection to use the method of ramping the bias current, for each event occurs in a time span $1/v$, and the performances improve as the speed is increased.

Future research could be directed to analyze the detection performances of a Josephson junction series array, possibly coupled to a resonator [41--44], or to investigate the effect of non-Gaussian noise [35,36,45]. Also, having limited the exploration of the JJ parameters to the case $\alpha = 0.2$ and $D = 0.05$, our results should be regarded as indicative, rather than an experiment design. More values of the parameters, especially noise amplitude, should be considered, possibly performing the simulations with CUDA [40].

\begin{acknowledgments}
The authors thank C.~Barone, C.~Guarcello, and S.~Pagano for useful discussions.
\end{acknowledgments}

\end{document}